\documentstyle[preprint,aps]{revtex} % for preprint
\preprint{          Preprint numbers: \ \
          UUHEP 93/3 \ \
          IUHET--259 \ \
          AZPH--TH/93--28 \ \
          UCSBTH--93--19
}
\begin{document}
%\draft
%
%%%%%%%%%%%%%%%%%%%%%%%%%%%%%%%%%%%%%%%%%%%%%%%%%%%%%%%%%%%%%%%%%%%%%%%%
%
\title{
Baryon Density Correlations\\ in High Temperature Hadronic Matter
}
\author{
Claude Bernard
}
\address{
Department of Physics, Washington University, St.~Louis, MO 63130, USA
}
\author{
Thomas A.~DeGrand
}
\address{
Physics Department, University of Colorado, Boulder, CO 80309, USA
}
\author{
Carleton DeTar
}
\address{
Department of Physics, University of Utah, Salt Lake City, UT 84112
}
\author{
Steven Gottlieb
}
\address{
Department of Physics, Indiana University, Bloomington, IN 47405, USA
}
\author{
Alex Krasnitz
}
\address{
IPS, RZ F 3, ETH-Zentrum, CH-8092 Zurich, SWITZERLAND
}
\author{
Robert L.~Sugar
}
\address{
University of California, Santa Barbara, CA 93106, USA
}
\author{
Douglas Toussaint
}
\address{
Department of Physics, University of Arizona,
Tucson, AZ 85721, USA
}
%
%\receipt{}
%
\date{\today}
\maketitle
\newpage
\begin{abstract}
      As part of an ongoing effort to characterize the high
temperature phase of QCD, in a numerical simulation using the
staggered fermion scheme, we measure the quark baryon density in the
vicinity of a fixed test quark at high temperature and compare it with
similar measurements at low temperature and at the crossover
temperature.  We find an extremely weak correlation at high
temperature, suggesting that small color singlet clusters are
unimportant in the thermal ensemble.  We also find that at $T = 0.75\
T_c$ the total induced quark number shows a surprisingly large
component attributable to baryonic screening.  A companion simulation
of a simple flux tube model produces similar results and also suggests
a plausible phenomenological scenario: As the crossover temperature is
approached from below, baryonic states proliferate.  Above the
crossover temperature the mean size of color singlet clusters grows
explosively, resulting in an effective electrostatic deconfinement.
\end{abstract}
%
%%%%%%%%%%%%%%%%%%%%%%%%%%%%%%%%%%%%%%%%%%%%%%%%%%%%%%%%%%%%%%%%%%%%%%%
\section{Introduction}

Numerical simulations of the quark plasma have suggested seemingly
contradictory models.  While bulk thermodynamic quantities, such as
the energy density\cite{ref:endens} and baryon
susceptibility\cite{ref:barsus} yield values consistent with a nearly
free gas of quarks and gluons, measurements of screening propagators,
particularly, measurements of the wave functions of exchanged objects,
are consistent with the confinement of color
singlets\cite{ref:screen_wavef}. Indeed, simulations and analytic work
in the pure glue sector have demonstrated that space-like Wilson loops
obey an area law in the high temperature phase, a signature of
confinement\cite{ref:Borgs_PM}.

One resolution of this seeming paradox describes the quark plasma as
an ensemble of color singlet clusters of various sizes.  Bulk
thermodynamic quantities, such as the energy density, would receive
contributions from all clusters, whereas long-range screening would be
controlled by the lightest clusters.  How large is the typical color
singlet cluster?  What is the typical spatial extent and quark and
antiquark content?  To answer these questions, it is necessary to seek
observables that have not hitherto been studied in this context.
Thus, we measured the distribution of induced quark charge (baryon
number) in the vicinity of a fixed test quark, at low and high
temperature, and at the crossover temperature.  This observable has
also been studied by the Vienna group in an effort to discern changes
in the QCD vacuum induced by color charges\cite{ref:FFM}. At low
temperature we expect that, as a result of confinement, a dynamical
antiquark or, less often, a pair of quarks, screens the test charge at
short distance.  Thus, the induced dynamical quark number density
should be large and negative close to the test charge.  If screening
is entirely due to a single antiquark, we should observe that the
total induced quark number $Q$ is $-1$.  By contrast, if color singlet
clusters are large either in size or in the number of quarks and
antiquarks, we would expect only a small induced charge density near
the source.

In Sec.~\ref{sec:theory} we describe the observable in detail and in
Sec.~\ref{sec:results} we present the results of the numerical
simulation.  Among the more striking results is the surprising
weakness of the induced quark number density at high temperature.  We
also find that at temperatures near, but below the crossover, the
total induced charge is significantly different from what would be
expected if only a single antiquark were responsible for
screening.\cite{ref:lattice93}

To help understand these results we turned to a simple flux tube model
of Patel\cite{ref:Patel}.  This model incorporates some of the
essential features of QCD, but has the added appeal that it can be
formulated simply, either in a field-diagonal basis analogous to the
Wilson QCD action or in a phenomenologically suggestive flux-diagonal
basis.  The latter representation permits the study of the growth and
complexity of the color singlet clusters of the model.  The induced
quark density and charge can also be studied in this model.  Results
of this study are presented in Sec.~\ref{sec:fluxtube}.  We find
results surprisingly similar to those for QCD\@.  A direct examination
of color singlet cluster size in this model suggests an explanation
for the QCD results, including a description of the nature of the
phase transition, and of the structure of the high temperature phase.
In particular, it suggests that heating of the confined phase results
in the appearance of quark clusters, including a surprisingly high
number of baryons and antibaryons.  As the crossover temperature is
passed, these clusters grow explosively, both in size and in quark
content, resulting in a suppression of baryonic correlations, a rise
in the baryon susceptibility, and an effective electrostatic
deconfinement.

A concluding discussion is given in Sec.~\ref{sec:conclusion}.

\section{Induced baryon density for staggered fermions}
\label{sec:theory}

Here we derive the staggered fermion observables to be measured in the
numerical simulation.

\subsection{Baryon density from the local chemical potential}

The construction of the local quark number density starts with the
introduction of a baryon chemical potential in the standard
way\cite{ref:chem_pot}, but with a spatial dependence.
Such a definition assures that the total baryon charge so defined is
exactly conserved on the lattice.  We start with the lattice action
for staggered fermions in the notation of Ref.\ \cite{ref:barsus},
modified through the introduction of a {\em local} chemical potential.
Let $r = ({\bf r},t)$ and $r^\prime = ({\bf r^\prime},t^\prime)$ denote
lattice coordinates.  The action is
\begin{equation}
   S(U,\psi,\bar\psi) = S_g(U) +
   \sum_{\bf r,\bf r^\prime}\bar\psi(\bf r)M(U)_{r,r^\prime}\psi(\bf r^\prime)
\end{equation}
where $S_g(U)$ is the pure gauge action.  The fermion flavor is
implicitly summed over. The fermion matrix for a single flavor with a
local chemical potential $\mu({\bf r)}$ is given by
\begin{equation}
	M(U)_{r,r^\prime} = 2ma\delta_{r,r^\prime} + 	\sum_{\nu \in
        \{\hat x,\hat y,\hat z\}}
	[\tilde U_{r,\nu} \delta_{r,r^\prime-\nu}
                    - \tilde U^\dagger_{r-\nu,\nu} \delta_{r,r^\prime+\nu}]
        + M^t(U)_{r,r^\prime}
\end{equation}
where the link matrices include the usual Dirac phase factor
\begin{equation}
  \tilde U_{r,\nu} = \eta_{r,\nu}U_{r,\nu}
\end{equation}
The Dirac phase factors $\eta_{r,t}$ also include the sign for the
antiperiodic boundary condition.
We consider two alternative formulations ``static'' and
``slice'' for the the time-hopping part of the fermion matrix $M^t$.  First
\begin{equation}
M^t_{\rm static}(U)_{r,r^\prime} =
 [\tilde U_{r,t}e^{a\mu({\bf r)}}
 \delta_{r,r^\prime-\hat t}
 - \tilde U^\dagger_{r-t,\hat t}e^{-a\mu({\bf r})}
 \delta_{r,r^\prime+\hat t}]
\label{eq:static}
\end{equation}
Notice that the fugacity factor has
been spread uniformly in the time dimension.  We also consider an
alternative ``single-slice'' definition that introduces the fugacity
factor on a single time slice:
\begin{eqnarray}
M^t_{\rm slice}(U)_{r,r^\prime} &=&
 [\tilde U_{r,t}e^{N_ta\mu({\bf r)}}
 \delta_{r,r^\prime-\hat t}\delta_{t^\prime,0}
 - \tilde U^\dagger_{r-t,\hat t}e^{-N_ta\mu({\bf r})}
 \delta_{r,r^\prime+\hat t}\delta_{t,0}]\\ \label{eq:single_slice}
	&&+
 [\tilde U_{r,t}\delta_{r,r^\prime-\hat t}(1 - \delta_{t^\prime,0})
   - \tilde U^\dagger_{r-\hat t,t  }\delta_{r,r^\prime+\hat t}(1 -
\delta_{t,0})]\nonumber
\end{eqnarray}
Of course, which formulation we choose depends on what we want to
measure.  If the chemical potential $\mu({\bf r})$ is introduced on a
single time slice as in Eq.~(5),
%Eq.~(\ref{eq:single_slice})  %% This didn't work.  A bug in ReVTeX???
it is a source
for the local baryon density at a single time.  If it is instead
spread over all time as in Eq.~(\ref{eq:static}), it is a source for
the time-averaged (static) baryon density.  When $\mu({\bf r})$ is
independent of ${\bf r}$ it can be shown that the determinant of the
fermion matrix is the same in either case.  Thus, for example, the
baryon susceptibility\cite{ref:barsus} is obtained by differentiating
the free energy twice with respect to such a constant chemical
potential, so it is the same in either formulation.  Moreover, for any
local time-independent observable, such as the baryon density in the
presence of a static charge, the expectation value is the same.  A
difference appears in the dynamical baryon density-density
correlation, which may depend on the time separation of the operators.
The single-slice formulation gives the equal-time density-density
correlation, and the static formulation gives the correlation of the
time-averaged densities.  For those observables that are independent
of the formulation, we find that the static version has the practical
advantage that, because of time averaging, it produces a higher signal
to noise ratio.

The baryon density per unit lattice cell $a^3$ at zero chemical
potential at a chosen spatial coordinate {\bf r} is given in terms of
the individual flavor densities by
\begin{equation}
  \rho({\bf r}) = \sum_{i=1}^{N_f} \rho_i({\bf r})
\end{equation}
where
\begin{equation}
     \rho_i({\bf r}) = \beta^{-1}
	\partial \log Z/ \partial \mu_i({\bf r})|_{\mu_i({\bf r})=0}.
\end{equation}
Here $\beta = 1/T$.
Now, for two flavors of staggered fermions the partition function is
\begin{equation}
   Z = \int[dU]\exp[-S_g(U)] [\det M_u(U,\mu_u) \det M_d(U,\mu_d)]^{1/4}
\end{equation}
and we have
\begin{equation}
       \left\langle \rho_{u,d}({\bf r})\right\rangle
       =(4\beta)^{-1}\left\langle {\rm Tr}[M_{u,d}^{-1}
\partial M_{u,d}/\partial\mu_{u,d}({\bf r})]\right\rangle_U
|_{\mu_{u,d}({\bf r})=0}
\end{equation}
The result is
\begin{equation}
  \left\langle \rho_{u,d}({\bf r})\right\rangle = (1/4)\phi_\tau\left\langle
 {\rm Tr}_c[M^{-1}_{({\bf r},\tau+1),({\bf r},\tau)} \tilde U_{({\bf
r},\tau);t}]
 +  {\rm Tr}_c[M^{-1}_{({\bf r},\tau),({\bf r},\tau+1)}
 \tilde U^\dagger_{({\bf r},\tau);t}]\right\rangle_U
\end{equation}
where $\tau$ is summed over and ${\rm Tr}_c$ denotes a trace over
color indices only.  As a device for treating both cases ``static''
and ``slice'' for the time-hopping term in the same expression, we
have introduced the weight
\begin{equation}
\phi_{\rm static,\tau} = 1/N_t \ \ \ \ \ \phi_{\rm slice,\tau} =
\delta_{\tau,0}.
\end{equation}
Because the hopping term in the staggered fermion matrix has the
time-reversal symmetry
\begin{equation}
M^{-1}_{r,r^\prime} = (-)^{r-r^\prime}M^{-1\dagger}_{r^\prime,r}
\label{eq:antih}
\end{equation}
the two terms on the rhs of the expression for the density are
negative complex conjugates of each other.  We get
\begin{equation}
       \left\langle\rho_{u,d}({\bf r})\right\rangle =
       (i/2){\rm Im}\left\langle P_{\rm dyn}({\bf r})\right\rangle
\label{eq:rho_pdyn}
\end{equation}
where
\begin{equation}
 P_{\rm dyn}({\bf r}) =
	{\rm Tr}_c\left[M^{-1}_{({\bf r},\tau+1),({\bf r},\tau)}\tilde U_{({\bf
r},\tau);t}\phi_\tau\right].
\end{equation}
Clearly for the particular time-independent case of
Eq.~(\ref{eq:rho_pdyn}) both the ``single slice'' and ``static''
formulations yield the same result.  In particular, since the
expectation value on the rhs is related to the Polyakov loop
expectation value, which is real for the usual action at zero chemical
potential, we get zero for the baryon density at zero chemical
potential, as we should.

\subsection{Density in the presence of a test quark}

Now we want to consider the correlation between a point test quark at
the origin (say) and the baryon density at {\bf r}.  Introducing the
test quark simply involves modifying the action by including a
Polyakov loop factor $P_{\rm fixed}(0)$, defined through
\begin{equation}
  P_{\rm fixed}({\bf r}) = {\rm Tr}_c\left[\prod_{t=0}^{N_t}
	\tilde U_{({\bf r},t);t}\right].
\end{equation}
We then have the partition function for the ensemble with the test quark:
\begin{equation}
   Z_q = \int[dU]\exp[-S_g(U)]
   [\det M_u(U,\mu_u) \det M_d(U,\mu_d)]^{1/4} P_{\rm fixed}(0)
\end{equation}
The baryon density $\rho_q({\bf r})$ in the presence of the fixed
quark can be calculated on the original ensemble $Z$ through
\begin{equation}
 \rho_q({\bf r}) = (iN_f/2)
  \left\langle P_{\rm fixed}(0)
  {\rm Im} P_{\rm dyn}({\bf r})\right\rangle_U/
  \left\langle P_{\rm fixed}(0)\right\rangle_U.
\end{equation}
Now the expectation values are taken with respect to the original
test-charge-free ensemble.  The correlation between ${\rm Re} P_{\rm fixed}$
and
${\rm Im} P_{\rm dyn}$ vanishes because of the complex conjugation symmetry of
the integral over the gauge variables.  What survives is the
correlation of the imaginary parts divided by the expectation of the
real part of the Polyakov loop, namely
\begin{equation}
 \rho_q({\bf r}) = -(N_f/2)
	\left\langle{\rm Im} P_{\rm fixed}({\bf 0}){\rm Im} P_{\rm dyn}
   ({\bf r})\right\rangle_U/\left\langle{\rm Re}
   P_{\rm fixed}({\bf 0})\right\rangle_U.
\label{eq:density}
\end{equation}

\subsection{Density-density correlation}
The connected density-density correlation for flavors $i$ and
$j$ is given by
\begin{eqnarray}
 \rho_{ij}({\bf r}) &=& \left\langle\rho_i({\bf r})
  \rho_j({\bf 0})\right\rangle -
  \left\langle\rho_i({\bf 0})\right\rangle
  \left\langle\rho_j({\bf 0})\right\rangle \nonumber\\
  &=& \beta^{-2}
\partial^2 \log Z/ \partial \mu_i({\bf r},0)\partial \mu_j({\bf
0},0)|_{\mu_i({\bf r})=0}
\end{eqnarray}
In terms of the integrated correlation
\begin{equation}
 \rho_{{\rm tot},ij} = \int d^3r \rho_{ij}({\bf r})
\end{equation}
the baryon singlet and nonsinglet susceptibilities\cite{ref:barsus} are
\begin{eqnarray}
 \chi_S &=& \beta (\rho_{{\rm tot},uu} + \rho_{{\rm tot},dd} +
 \rho_{{\rm tot},ud} + \rho_{{\rm tot},du}) \\
 \chi_{NS} &=& \beta (\rho_{{\rm tot},uu} + \rho_{{\rm tot},dd} -
 \rho_{{\rm tot},ud} - \rho_{{\rm tot},du})
\label{eq:suscept}
\end{eqnarray}
In terms of the fermion matrices for the separate flavors the
correlation receives four contributions
\begin{equation}
\rho_{ij}({\bf r}) = S_{11ij} + S_{21ij} + S_{22ij} - S_{{\rm disc},ij},
\end{equation}
The indices $nm$ in $S_{nmij}$ count the number of explicit coordinate points
$n$ and number of color traces $m$.
\begin{eqnarray}
  S_{11ij} &=& (2\beta)^{-2}\delta_{ij}\left\langle
  {\rm Tr}[M_i^{-1}\partial^2 M_i/\partial\mu_i({\bf r})
  \partial\mu_i({\bf 0})]
  \right\rangle_U
  |_{\mu_{u,d}({\bf r})=0}
 \\
 S_{21ij} &=& -(2\beta)^{-2}\delta_{ij}\left\langle
 {\rm Tr}[M_i^{-1}\partial M_i/\partial\mu_i({\bf r})
 M_i^{-1}\partial M_i/\partial \mu_i({\bf 0})]
 \right\rangle_U|_{\mu_{u,d}({\bf r})=0}\\
 S_{22ij} &=& (4\beta)^{-2}\left\langle
 {\rm Tr}[M_i^{-1}\partial M_i/\partial\mu_i({\bf r})]{\rm Tr}[
 M_j^{-1}\partial M_j/\partial \mu_j({\bf 0})]\right\rangle_U
 |_{\mu_{u,d}({\bf r})=0}\\
 S_{{\rm disc}ij} &=& \left\langle\rho_i({\bf 0})\right\rangle
 \left\langle\rho_j({\bf 0})\right\rangle.
\end{eqnarray}
Carrying out the derivatives and using the time-reversal property
(\ref{eq:antih}) gives
\begin{eqnarray}
S_{11ij} &=& 2^{-1}\delta_{ij}\delta_{{\bf r},{\bf 0}}\left\langle
	{\rm Re} P_{\rm dyn}({\bf 0})\right\rangle \\
S_{21ij} &=& -2^{-1}\delta_{ij}\phi_{\tau^\prime}\phi_\tau\left\langle
	{\rm Re} {\rm Tr}_c[M^{-1}_{({\bf 0},\tau+1),({\bf r},\tau^\prime)}
        \tilde U_{({\bf r},\tau^\prime);t}
	M^{-1}_{({\bf r},\tau^\prime+1),({\bf 0},\tau)} \tilde
        U_{({\bf 0},\tau);t}]\right\rangle_U \nonumber\\
     &&-2^{-1}\delta_{ij}\phi_{\tau^\prime}\phi_\tau\left\langle
        {\rm Re} {\rm Tr}_c[M^{-1}_{({\bf 0},\tau+1),({\bf r},\tau^\prime+1)}
        \tilde U^\dagger_{({\bf r},\tau^\prime);t}
	M^{-1}_{({\bf r},\tau^\prime),({\bf 0},\tau)} \tilde U_{({\bf
        0},\tau);t}]\right\rangle_U \\
S_{22ij} &=& -4^{-1}\left\langle{\rm Im} P_{\rm dyn}({\bf r}) {\rm Im} P_{\rm
dyn}({\bf 0})\right\rangle_U
\end{eqnarray}
The two-point single-trace term $S_{21ij}$ is a hadron propagator with a
source at $({\bf 0},\tau)$ and a sink at $({\bf r},\tau^\prime)$.  The source
and
sink are both just the point-split baryon density operator.  To
evaluate this term we require the quark propagator from the source
$\tilde U_{({\bf 0},\tau);t}$ at $({\bf 0},\tau)$ and the antiquark propagator
from
a point source at $({\bf 0},\tau+1)$.  These propagators are combined in
two ways at $({\bf r},\tau^\prime)$ to complete the evaluation of the two
contributions.

To evaluate the two-point, two-trace term, $S_{22ij}$ we use the random
source trick of Ref.\ \cite{ref:barsus}.  We introduce a set of $n_{\rm rand}$
independent complex Gaussian random SU(3) vectors $R_\ell({\bf r},\tau)$
$\ell=1,\ldots{},n_{\rm rand}$ on
the time slice $\tau=1$ for the ``slice'' form of the action and for all
$\tau$ for the ``static'' form.  Then on a
given gauge configuration $U$ we generate the Fourier transform of the
estimate of ${\rm Im} P_{\rm dyn}({\bf r})$.
\begin{equation}
I_{{\rm dyn},\ell}({\bf k}) = \sum_{\bf r} \exp(i{\bf r}\cdot{\bf k})\phi_\tau
{\rm Im}[ R_\ell({\bf r})^*
M^{-1}_{({\bf r},\tau+1),({\bf r},\tau)}\tilde U_{({\bf r},\tau),t}
R_\ell({\bf r})].
\end{equation}
(Note that the imaginary part is taken before carrying out the Fourier
transform.)

Finally, we estimate the Fourier transform of $S_{22ij}$ from
\begin{equation}
S_{22ij}({\bf k}) = -\frac{1}{4n_{\rm rand}(n_{\rm rand}-1)}\sum_{\ell \ne
\ell^\prime}
\left\langle I^*_{dyn,\ell}({\bf k}) I_{dyn,\ell^\prime}({\bf k})
\right\rangle_U.
\end{equation}
The single-point, single-trace term can be generated trivially from
the random source method through the estimate
\begin{equation}
S_{11ij} = \frac{\delta_{ij}\delta_{{\bf r},{\bf 0}}}{2n_{\rm rand}V}
\sum_\ell\sum_{\bf r}
{\rm Re}[R_\ell({\bf r},\tau)^*M^{-1}_{({\bf r},\tau+1),({\bf r},\tau)}
\tilde U_{({\bf r},\tau),t} R_\ell({\bf r},\tau)].
\end{equation}
Thus the computation of $S_{11ij}$ and $S_{22ij}$ starts with the
evaluation of the quark propagator from the same parallel-transported
random source, namely, $\tilde U_{({\bf r},\tau),t}R_k({\bf r},\tau)$.
The same Fourier transform $I_{dyn,i}$ is used to obtain the
correlation with the static source.  However, evidently the
computation of the hadron propagator $S_{21ij}$ must be done with
point sources---not random sources.

\section{Results of simulations}
\label{sec:results}

Simulations were carried out at fixed $\beta = 5.445$ and quark mass
$am_q = 0.025$ for two flavors of staggered fermions on lattices of
size $16^3 \times N_t $, where $N_t = 8,6,4$.  This choice of lattice
parameters corresponds to the crossover temperature at $N_t = 6
$\cite{ref:Nt6}.  Thus, the simulations are done at three temperatures
$T = 0.75\ T_c $, $T \approx T_c$, and $T = 1.5\ T_c$, respectively, at
the {\em same} lattice scale, making it meaningful to
superimpose plots of baryon density {\it vs} distance from the source.
Spectroscopic simulations at the same temperature\cite{ref:spectrum}
allow us to set the scale, {\it viz.} $T_c = 145 $ MeV and $a = 0.227 $ fm.
Simulations were also carried out at $N_t = 4$ with $\beta = 5.15$,
$5.22$, $5.25$ and $5.29$ (the crossover) to provide an independent
check of trends in the total induced quark number.  Table~\ref{tab:simsize}
shows the extent of the simulation sample.

Figure~\ref{fig:dens3T} summarizes our results for the induced quark
number density at these three temperatures.  Particularly striking is
the dramatic decrease in the correlation at high temperature.  Thus,
we see no evidence for small color singlet clusters in the high
temperature plasma.

The total induced quark number normalized to one for a single quark was
computed in two ways: first by a direct integration of the density
(\ref{eq:density}) and second by fitting the density distribution to
the functional form
\begin{equation}
   \rho_q({\bf r}) = \sum_j \sum_{\bf k} e^{2\pi i {\bf k}\cdot {\bf r}/N}
            \frac{a_j}{\sum_{i=1}^3 2\cos(2\pi k_i/N) + m_j^2 - 6},
\end{equation}
(the sum of free lattice propagators for scalar fields of
mass $m_j$) from which the total induced quark number is
\begin{equation}
   Q = N^3 \sum_j a_j/m_j^2 .
\end{equation}
At most two mass terms were used.  All fits started at zero radius.
The resulting total quark number determined from the two methods was
in each case consistent within errors.  The value quoted is the one
with the smaller standard deviation.  It was found that except for the
high temperature points ($6/g^2 = 5.445$, $N_t = 4$ and $6/g^2 =
5.29$, $N_t = 6$) direct summation over the entire volume gave poorer
statistics than fitting, because direct summation suffers from
statistical noise introduced by contributions far from the fixed
charge that should sum to zero.  By fitting to a functional form that
falls to zero at infinity, we controlled this noise.

The resulting fitted curves are plotted in Fig.~\ref{fig:dens3T}.  The
total quark number values are also given in the legend and in Table
\ref{tab:scalar5445}.  At the high temperature point the total induced
quark number is nearly two orders of magnitude smaller that the quark
number at low temperature.  At low temperature we expect that the test
charge is attached to a color singlet cluster.  A single antiquark
would contribute $-1$ to the total induced quark number, and a pair of
quarks forming a baryon, $+2$.  A thermodynamic mixture of these two
configurations would give an intermediate value.  Based on the error
ellipse for a two-parameter fit to the lowest-temperature density, at
the two standard deviation level, we find that the total induced quark
number is greater than $-0.55$, significantly different from $-1$.

As a check of this result, we also carried out a series of simulations
at varying $\beta$ with $N_t = 4$.  At such a strong coupling the
Polyakov loop expectation value is large, leading to a stronger signal
in the correlation.  Results for the induced quark number are plotted
in Fig.~\ref{fig:q_vs_beta_nt4} and listed in
Table~\ref{tab:scalarnt4}.  The low temperature values are somewhat
larger in magnitude that in the $6/g^2 = 5.445$ simulation, but still
show a significant departure from $-1$.

For the sake of comparison let us estimate the contribution to the
induced charge from the lowest S-wave mesonic and baryonic screening
clusters in the ensemble.  These clusters are obtained by replacing
one quark in the $\pi$, $\rho$, $N$, and $\Delta$ by a fixed spinless,
flavorless color triplet quark.  The result is a modified $J=1/2$,
$I=1/2$ meson with a four-fold multiplicity $g_{\rho^\prime} = 4$, a
modified $J=0$, $I=0$ ``nucleon'' $N^\prime$ and a modified $J=1$,
$I=1$ delta $\Delta^\prime$.  In the continuum limit the degeneracies
are $g_{N^\prime} = 1$ and $g_{\Delta^\prime} = 9$.  However, at
finite lattice spacing the $\Delta^\prime$ is split, owing to the
discrete nature of the internal symmetries in the staggered fermion
scheme.  We have not measured masses of the modified states, but have
masses for their light quark counterparts\cite{ref:spectrum} for the
same coupling $6/g^2 = 5.445$, namely, $am_\rho = 0.918$, $am_\pi =
0.4488$, $am_N = 1.375$, and $am_\Delta = 1.43$.  If we assume that
the splitting of the $\rho$ and $\pi$ is entirely due to the color
hyperfine interaction, then we estimate the mass of the modified meson
to be $m_{\rho^\prime} = M + (3m_\rho + m_\pi)/4 = M + 0.80/a$, where
$M$ represents the contribution from the point charge.  Similarly, we
have $m_{N^\prime} = M + m_N = M + 1.375/a$ and $m_{\Delta^\prime} = M +
(2m_\Delta + m_N)/3 = M + 1.41/a$.  To be conservative, let us assume that
the $\Delta^\prime$ is fully degenerate.  As noted before, the induced
quark number is
\begin{equation}
   Q = -p_m + 2p_b,
\end{equation}
where $p_m$ and $p_b$ are the probabilities of screening via the
mesonic and baryonic clusters.  For this estimate we take $p_m + p_b =
1$.  The probabilities are estimated from the Boltzmann weights:
\begin{equation}
   p_m/p_b = \frac{g_{\Delta^\prime} e^{-m_{\Delta^\prime}/T} +
      g_{N^\prime}e^{-m_{N^\prime}/T}}{g_{\rho^\prime} e^{-m_{\rho^\prime}/T}}
     = \frac{9e^{-1.41/aT} + e^{-1.375/aT}}{4e^{-0.80/aT}}
\label{eq:relwt}
\end{equation}
The unknown regularization-dependent energy $M$ has cancelled in the
ratio.  The resulting estimates are $Q = -0.94$ at $T = 0.75\ T_c =
1/8a$ and $Q = -0.81$ at $T = T_c = 1/6a$.  These values are
considerably lower than were found in the simulation.  To bring the
estimates into closer agreement would require adding more baryonic
states.  Thus the full simulation suggests that already at a
temperature of $0.75\ T_c$, there is significant baryonic screening of
the fixed charge.

As we have remarked (\ref{eq:suscept}), the integral of the
self-correlation of the dynamical quark number density gives the
baryon susceptibility.  Results for the susceptibility on $8^3 \times
4$ lattices with $ma = 0.025$ were reported in Ref.~\cite{ref:barsus}.
The high temperature values found are consistent with what would be
expected for an ideal quark gas, given the very large effects of the
nonzero lattice spacing.  As a check, we compare our new results on
$16^3 \times 4$.  Figures \ref{fig:xns_nt4} and \ref{fig:xs_nt4} show
the comparison for the $N_t = 4$ series.  They are apparently
consistent.  Results for the $N_t = 4$ series are also given in Table
\ref{tab:scalarnt4} and for the $6/g^2 = 5.445$ series, in Table
\ref{tab:scalar5445}.  Not surprisingly, the static form of the
density operator gives a better signal than the single time slice
form, because it involves an average over all time links.  Values are
not available for $\beta = 5.29$, which was run before improvements in
the code incorporated the static form and gave an acceptable signal to
noise ratio for the slice form.

The correlation between a test charge and the scalar density
$\left\langle\bar\psi \psi\right\rangle$ has been measured by the Vienna
group\cite{ref:FFM}.  Our results, shown in Fig.~\ref{fig:pbp_vs_r},
are consistent with theirs.

\section{Flux tube model}
\label{sec:fluxtube}

\subsection{The model}

Some years ago Patel\cite{ref:Patel} proposed a flux tube version
of the three-state three-dimensional Potts model to explain the
mechanism of the deconfining phase transition in QCD\@.  In this
model, each site $\bf r$ of a cubic lattice holds either a quark,
antiquark, or none at all, and each link $\ell_{\bf r,\mu}$, a triplet
or antitriplet flux, or none at all.  That is the quark number $n_{\bf
r}$ and the flux $\ell_{\bf r,\mu}$ take on values $\{-1,0,1\}$. Flux
is conserved modulo 3.
\begin{equation}
  \sum_{\mu=1}^3 \ell_{\bf r,\mu} + \ell_{\bf r,-\mu} - n_{\bf r} = 0 \bmod 3
   \label{eq:gauss}
\end{equation}
where $\ell_{\bf r,-\mu} = -\ell_{\bf r-\hat\mu,\mu}$. The hamiltonian
is given in terms of the quark mass $m$ and the string link energy
$\sigma$ by
\begin{equation}
   H = \sum_{\bf r,\mu} \sigma |\ell_{\bf r,\mu}| + \sum_{\bf r}m |n_{\bf r}|.
\end{equation}

The partition function is then
\begin{equation}
  Z(\beta) = \sum_{\{\ell_{\bf r,\mu},n_{\bf r}\}^\prime} \exp(-\beta H)
\end{equation}
where the prime signifies a sum constrained by (\ref{eq:gauss}).

\subsection{Equivalence to the Potts model}

This model is equivalent to the three-state three-dimensional Potts
model.\cite{ref:Patel}  The equivalence can be seen by replacing the Gauss' law
constraint by
\begin{equation}
   \delta_{\ell,0} = \frac{1}{3} \sum_{z \in Z(3)} z^\ell
\end{equation}
in modulo three arithmetic.  Here $Z(3) = \{1,e^{\pm 2\pi/3}\}$.
Introducing this identity on each site with the summation variable
$z_{\bf r}$ allows us to rewrite the partition function as
\begin{equation}
  Z(\beta) = \sum_{\{\ell_{\bf r,\mu},n_{\bf r}, z_{\bf r}\}}
  \exp\left(-\beta \left[\sum_{\bf r,\mu} \sigma |\ell_{\bf r,\mu}| +
  \sum_{\bf r} m |n_{\bf r}|\right]\right)
   \prod_{\bf r} z_{\bf r}^{\sum_{\mu}(\ell_{\bf r,\mu} +
  \ell_{\bf r,-\mu}) - n_{\bf r}}
\end{equation}
The unconstrained sums over links and quark numbers can be carried out
as follows:
\begin{eqnarray}
  \sum_{\ell_{\bf r,\mu}}\exp(-\beta|\ell_{\bf r,\mu}|)
   ( z_{\bf r} z_{\bf r+\hat \mu}^*)^{\ell_{\bf r,\mu}} &=&
   1 + 2{\rm Re}(z_{\bf r} z_{\bf r+\hat \mu}^*)\exp(-\beta\sigma) \\
   \sum_{n_{\bf r}}\exp(-\beta m |n_{\bf r}|)z_{\bf r}^{-n_{\bf r}}
   &=& 1 + 2{\rm Re} z_{\bf r}.
\end{eqnarray}
Thus the partition function reduces to a product of polynomials in the
Z(3) variables $z_{\bf r}$.  A hamiltonian can be constructed through
the identity over Z(3)
\begin{equation}
  \log(1 + cz + cz^*) = a + bz + bz^*
\end{equation}
where
\begin{eqnarray}
  \exp(2a) &=& (1+2c)(1-c)^2 \\ \nonumber
  \exp(3b) &=& (1+2c)/(1-c)
\end{eqnarray}
Thus if we define
\begin{eqnarray}
  J\beta^\prime &=&
  \frac{2}{3}\ln\left(\frac{1+2\exp(-\beta\sigma)}{1-\exp(-\beta\sigma)}\right)
\\
  h\beta^\prime &=&
  \frac{2}{3}\ln\left(\frac{1+2\exp(-\beta m)}{1-\exp(-\beta m)}\right),
  \label{eq:fluxtoPotts}
\end{eqnarray}
then we have, up to a constant
\begin{equation}
  Z(\beta) = \sum_{\{z_{\bf r}\}}\exp(-\beta^\prime H^\prime)
\end{equation}
where
\begin{equation}
  H^\prime = -\sum_{\bf r\mu}J {\rm Re} (z_{\bf r} z_{\bf r+\hat \mu}^*)
  - \sum_{\bf r} h {\rm Re} z_{\bf r}.
\end{equation}
This expression is recognized as the hamiltonian of the three-state
Potts model in three dimensions with a coupling $J$ and a magnetic
field $h$ coupled to the real part of the spin.

Thus a low quark mass corresponds to a high Potts magnetic field and
a low flux-model temperature corresponds to a high Potts temperature.
At zero field $h$ a first order phase transition is found in this
model at $J\beta^\prime \approx 0.367$.  The phase transition persists
for a small magnetic field $h\beta^\prime < 0.002$, but is not evident
in numerical simulations for larger values of the field\cite{ref:DD}.
These parameter values can be converted to the flux tube model values
through the inverse of Eq.\ (\ref{eq:fluxtoPotts}):
\begin{eqnarray}
  \beta \sigma &=& \ln\left(\frac{\exp(\frac{3}{2}J\beta^\prime)+2}
  {\exp(\frac{3}{2}J\beta^\prime)-1}\right) \\
  \beta m &=& \ln\left(\frac{\exp(\frac{3}{2}h\beta^\prime)+2}
  {\exp(\frac{3}{2}h\beta^\prime)-1}\right),
\end{eqnarray}
giving a phase transition along a curve starting at about $\beta\sigma
= 1.63$, $\beta m = \infty$ to about $\beta m > 6.9$ or $m/\sigma >
4.2$.

\subsection{Relationship to QCD}

Patel proposed using this model as a paradigm for the QCD phase
transition.  The Potts model in its more conventional form was also
offered some years ago as a model of the deconfining phase
transition\cite{ref:SY}. The latter formulation is obtained from a
high-quark-mass, strong-coupling, high-temperature, Z(3)-restricted
approximation to the conventional field-diagonal Wilson action, with
the Potts spin corresponding to the Polyakov loop.  On the other hand
the flux-tube formulation of the Potts model corresponds to an
alternate representation of the Wilson action in the
charge-and-flux-diagonal basis.  Because of the combinatoric
complexities of linking SU(3) charges and fluxes to form color
singlets, such a basis for the Wilson action never received wide
attention.  However, in the simple Z(3) basis of the flux tube model
the combinatorics become trivial.  Moreover, in the flux-tube form
the model offers the highly suggestive possibility of studying the
size and structure of color singlet clusters as a function of
temperature.  Its chief drawback is that, because it treats quarks as
static objects, it does not incorporate chiral symmetry.

At low temperature only small color singlet clusters populate the
Gibbs ensemble.  As the temperature is increased, clusters of
increasing size occur.  Eventually clusters connect to fill nearly
the entire spatial volume.  For heavy quark masses this phenomenon
leads to a first-order deconfinement phase transition.  For
light quarks, cluster growth is somewhat inhibited, since pair
formation breaks the flux links.  It is found that there is no phase
transition.  One is tempted to think of a percolation phase
transition mediated by the connectivity of the flux tubes, but there
is nothing to percolate: there is no current in the model to flow
between linked sites and establish long-range order.  The first-order
phase transition occurring only at large quark mass (low magnetic
field) resembles more appropriately a liquid-gas phase transition.

\subsection{Simulation and Observables}

To simulate the model in the flux tube basis, we used a Metropolis
algorithm, with moves designed to preserve Gauss' law.  We considered
two types of elementary moves: the addition (or removal) of the
lightest meson (flux link with quark and antiquark at the ends) and
the addition (or removal) of the lightest ``glueball'' (four flux
links directed around a plaquette).  Adding was done in the literal
sense: adding a quark to a site means increasing the quark number by
one unit, modulo three, etc.  Thus, the actual impact of these
elementary moves depends on the configuration.  The moves could
result in shortening, lengthening, breaking, joining, or displacing a
series of flux links.  Notice that two or three such mesons can form
a baryon, if they are added with the antiquark on the same site but
with links and quarks on unique sites, so we didn't create baryons
through a separate move.  Although by the same token four such mesons
can form a glueball, since meson formation is suppressed at high
quark mass, we kept both moves to allow a more efficient
mass-independent coverage of the ensemble.

A fixed charge is introduced at the origin by starting from a modified
vacuum configuration in which the dynamical quark charge at the origin
is set to $-1$ and all other charges and fluxes are initially zero.
Observables include these:
\begin{center}
  \begin{tabular}{ll}
  $|n|$ & mean number of quarks plus antiquarks per site \\
  $|\ell|$ & mean number of links (either sign) per link \\
  $n_{\rm vtx}$ & mean number of three-point flux vertices \\
  $\rho({\bf r})$ & induced quark number density \\
  $Q$ & the total induced charge (quarks minus antiquarks)\\
  $\left\langle B^2\right\rangle$ & mean square baryon number
  (including test charge) \\
  $N_0$ & the size of the cluster connected to the origin \\
  \end{tabular}
\end{center}

\subsection{Results}

Measurements were made on a $10^3$ lattice for a variety of $\beta$ at
$m = 1.0$ (10,000 sweeps for each $\beta$ value) and $m = 3.5$ (1,000
sweeps for each $\beta$ value).  We use units in which $\sigma = 1$.
As we have noted above, both quark masses are in a region where a
phase transition does not occur.  The higher mass series comes closer
to the critical point.  With such a small volume we would notice a
significant finite-size rounding of the first order phase transition
that occurs at higher quark masses.  However, we are primarily
interested in the qualitative behavior of the model at smaller quark
mass, since it corresponds more closely to the light quark QCD
simulation.  Thus the small volume suffices.

A popular indicator of the phase transition in QCD is the Polyakov
loop ${\rm Re} P$.  The Potts model analog is the magnetization,
\begin{equation}
\left\langle{\rm Re} z\right\rangle =
\frac{\partial \log Z}{\beta^\prime \partial h}.
\end{equation}
 From the map Eq~(\ref{eq:fluxtoPotts}) onto flux tube variables, we see
that the flux tube analog of this order parameter is the mean quark
count (quarks plus antiquarks)
\begin{equation}
\left\langle|n|\right\rangle = \frac{\partial \log Z}{V \beta \partial m}
\end{equation}
Figures \ref{fig:ftqk_en_10} and \ref{fig:ftqk_en_35} show the quark
number $|n|$ {\it vs} $\beta$ for the two quark masses.  The higher mass
series comes closer to the critical point, leading to a sharper
crossover.  The crossover locations are determined from the inflection
point of the curves (peak in the corresponding susceptibility) to be
$1/T_c = \beta_c = 1.88(1)$ for $m = 1.0$ and $\beta_c = 1.662(2)$ for
$m = 3.5$.  These values are used to determine the ratio $T/T_c$ in
Tables \ref{tab:ftscalars10} and \ref{tab:ftscalars35}.

Shown in Fig.~\ref{fig:ftqtot} and Tables \ref{tab:ftscalars10} and
\ref{tab:ftscalars35} is the total induced charge $Q$ as a function of
$\beta$ in the presence of a test charge for the two quark masses.
The results bear a striking resemblance to those of the QCD
simulation.  At high temperature, the induced charge is very small for
both quark masses.  For the heavier quark mass, the total induced
charge $Q = -0.75(7)$ at the highest beta ($T = 0.92\ T_c$) shows a
significant departure from $-1$.  For the lighter quark mass at the
highest $\beta$ ($T = 0.78\ T_c$) we also see a significant departure
with $Q = -0.69(2)$.

Thus we find evidence of important baryonic screening in this model at
temperatures close to the crossover.  Is this surprising?  Consider
the relative Boltzmann weights for the lightest baryon-like and
meson-like clusters attached to the test quark.  They are shown in
Fig.~\ref{fig:ftlightest}.  The masses are $m_m = M + m + \sigma$ and
$m_b = M + 2m + 2\sigma$ for the meson-like and baryon-like cluster,
respectively.  Here $M$ stands for the mass of the fixed charge.  For
$m = 1$ and $\sigma = 1$, corresponding to our the lighter quark mass,
these are $m_m = M + 2$ and $m_b = M + 4$.  The corresponding
multiplicities are $g_m = 6$ and $g_b = 30$.  At $\beta = 2.4$ we
have, in the notation of Eq (\ref{eq:relwt})
\begin{equation}
   p_b/p_m = 30e^{-4/T}/6e^{-2/T} = 0.04.
\end{equation}
Thus, if only these two states played a role in low temperature
screening we should find a total induced charge $Q = -0.88$,
significantly different from what is observed.  Clearly, still more
baryonic states contribute.  If we consider higher excitations, the
number of distinct baryonic states (i.e.\ states with total baryon
number $B \ne 0$) grows very rapidly with mass---much faster than
mesonic states.  Thus the entropy of baryon excitation increases the
importance of baryonic screening.  This point can be dramatized by a
direct examination of configurations.  Shown in
Fig.~\ref{fig:ftsample_lattice} is a representative configuration
obtained at the mass and inverse temperature in question: $m = 1$ and
$\beta = 2.4$.  It contains nine $q \bar q$ mesons, one $qqq$ baryon,
and one $\bar q \bar q \bar q$ antibaryon, a highly improbable
occurrence in the naive two-state model.

Turning now to the induced density {\it vs} distance, we show results for
the simulation at the lighter quark mass $m = 1.0$ in
Fig.~\ref{fig:ftdensityvsr}.  Again we see a resemblance to results of
the QCD simulation shown in Fig.~\ref{fig:dens3T}.  The total charge
$Q(r\le 3)$ given in the legend is found by integrating the density to
an arbitrary cutoff distance $r \le 3$.  The correlation is stronger
at low temperature than in the QCD simulation, reflecting a shorter
correlation length in lattice units, which could be adjusted by a
change of scale.  From a fit to a single-pole lattice Yukawa form, we
find an effective mass of 2.9(7) in the flux tube model at $\beta =
2.4$ and $m = 1.0$, to be compared with 1.7(6) for the corresponding
low temperature curve for SU(3) (Fig.~\ref{fig:dens3T}).

Further evidence of the importance of larger hadronic clusters can be
obtained from a measurement of the mean size of the cluster attached
to the origin.  This size is defined as the number of sites connected
through flux links to the origin.  Our results are shown in
Fig.~\ref{fig:ftclustersizevsT}.  Notice that in this $10^3$ lattice
the mean
cluster size grows dramatically, already filling 43\% of the volume at
$T \approx 1.2\ T_c$.  Despite appearances, however, correlation
lengths are nonetheless finite.

Another indicator of clustering is the number of three-point flux
vertices $n_{\rm vtx}$.  This statistic is the sum of one third the
total flux entering each site, if the total flux count at the site is
positive.  With coordination number six a site can contribute only 0,
1, or 2 to this statistic.  As can be seen from Tables
\ref{tab:ftscalars10} and \ref{tab:ftscalars35} this number grows with
temperature in much the same way for both the smaller and larger quark
mass.  On the other hand, a striking difference is seen in the
behavior of the baryon susceptibility $\left\langle
B^2\right\rangle /V$, as might be expected.  For the lighter quark mass
the fluctuation in baryon number is considerably larger at high
temperature.  The expected suppression of quark content with
increasing quark mass is also evident in the quark number per site
$|n|$.  An expected consequence of decreasing quark mass is a decrease
in cluster size, since pair creation breaks flux links.  This effect
is evident in the mean size of the cluster attached to the origin.  At
$m = 3.5$ and $\beta = 1.4$, corresponding to $T = 1.19\ T_c$, this
mean size 650(3).  When the quark mass is decreased to $m = 1.0$ at a
comparable temperature ($\beta = 1.6$), the mean size decreases to
430(3).  Thus, with decreasing quark mass the quark content of the
clusters increases, and the size decreases somewhat.

\section{Conclusions}
\label{sec:conclusion}

In numerical simulations we have measured the quark number density in
the vicinity of a test quark as a function of temperature.  A strong
correlation is found in the low temperature phase, but it is vastly
reduced in the high temperature phase.  We have also found evidence
for a significant proliferation of baryonic clusters as the crossover
temperature is approached from below.  A companion simulation of the
flux tube model has similar behavior and suggests an explanation.
Heating past the crossover in this model results in an explosive
growth of color singlet clusters.  Thus at high temperature the
addition of a single test quark has little effect on the ensemble,
leading to an extremely weak correlation.  We have an effective
electrostatic deconfinement without a phase transition.

We also find that in the flux tube model baryonic clusters proliferate
as the temperature rises through $T_c$, permitting more frequent
baryonic screening of a test charge, suggesting an explanation for an
apparent superabundance of baryons in full QCD at these temperatures.
It is interesting to speculate that such a superabundance,
particularly of antibaryons, should they survive final state
interactions, provides an experimental signal for the crossover to the
quark-plasma regime\cite{ref:antibaryon}.

To be sure the flux tube model omits many features of QCD\@.  It lacks
dynamics, describing only electrostatics.  It also ignores chiral
symmetry.  Completely omitted are the important magnetic interactions
that give rise to confinement in spacelike propagation.  It would be
useful to find an elaboration of the model more closely relevant to
QCD\@.  Nonetheless, it is highly suggestive both for further
exploration of QCD and for the phenomenology of the quark plasma.

\acknowledgments

Code development and testing were carried out on the nCUBE and Intel
iPSC/860 hypercubes at the San Diego Supercomputer Center.  The QCD
simulations were carried out on the Intel iPSC/860 at the NASA Ames
Research Center and on the Thinking Machines Corporation CM5 at the
National Center for Supercomputing Applications.  The flux tube
simulations were carried out on IBM RS6000/320 workstations in the
Physics Department of the University of Utah.  We wish to thank all of
these organizations for their support of our work. This research was
supported in part by Department of Energy grants
DE-2FG02--91ER--40628, %Claude
DE-AC02--84ER--40125, %Steve and Alex
DE-AC02--86ER--40253, %  Tom
DE-FG02--85ER--40213, %   Doug
DE-FG03--90ER--40546, %DOE iPSC/860 grant--really Julius's
DE-FG02--91ER--40661, %Steve's new grant
and National Science Foundation grants
NSF--PHY90--08482, NSF--PHY93--09458, %  Carleton
NSF--PHY91--16964 %  bob sugar
and NSF--PHY\-91--01853.  %NSF iPSC/860 grant
%
%%%%%%%%%%%%%%%%%%%%%%%%%%%%%%%%%%%%%%%%%%%%%%%%%%%%%%%%%%%%%%%%%%%%%%%%%%%%
%    References
%%%%%%%%%%%%%%%%%%%%%%%%%%%%%%%%%%%%%%%%%%%%%%%%%%%%%%%%%%%%%%%%%%%%%%%%%%%%

%%%%%%%%%%%%%%%%%%%%%%%%%%%%%%%%%%%%%%%%%%%%%%%%%%%%%%%%%%%%%%%%%%%%%%
%   Figure Captions
%%%%%%%%%%%%%%%%%%%%%%%%%%%%%%%%%%%%%%%%%%%%%%%%%%%%%%%%%%%%%%%%%%%%%%
%
\figure{
\caption{Quark number density induced by a fixed quark at the origin
as a function of distance from the origin at three temperatures.
Curves are fits to a single screening mass. The total induced quark
number $Q$ is also shown.}
\label{fig:dens3T}
% FIGURE_FILE fig01.ax
}
\figure{
\caption{Total induced quark number {\it vs} $6/g^2$ at $N_t = 4$. }
\label{fig:q_vs_beta_nt4}
% FIGURE_FILE q_vs_beta_nt4.ax = fig02.ax
}
\figure{
\caption{Nonsinglet quark number susceptibility {\it vs} $6/g^2$ at
$N_t = 4$ with bare quark mass $ma = 0.025$, compared with Ref.\
\protect\cite{ref:barsus} (crosses).  Two forms of the quark number density
are used: ``static'' (squares) and ``slice'' (circles).}
\label{fig:xns_nt4}
% FIGURE_FILE xs_vs_beta.ax = fig03.ax
}
\figure{
\caption{Singlet quark number susceptibility {\it vs} $6/g^2$ at
$N_t = 4$ with bare quark mass $ma = 0.025$.  Symbols are as in the
previous figure.}
\label{fig:xs_nt4}
% FIGURE_FILE xs_vs_beta.ax = fig04.ax
}
\figure{ \caption{Scalar density correlation $\left\langle\bar\psi
\psi\right\rangle$ {\it vs} distance from a point charge, normalized to one
at infinite distance.  Error bars for the two lowest temperature
points are as small or smaller than the plot symbol.  For the sake of
clarity, for the highest temperature only two typical error bars are
shown.} \label{fig:pbp_vs_r}
% FIGURE_FILE pbp.combo.ax = fig05.ax
}
\figure{ \caption{Total number of quarks and antiquarks {\it vs} inverse
temperature in the flux tube model for quark mass $m = 1.0$.  The
vertical line indicates the crossover.}
\label{fig:ftqk_en_10}
% FIGURE_FILE qe_vs_beta_10.ax = fig06.ax
}
\figure{\caption{Total number of quarks and antiquarks {\it vs} inverse
temperature in the flux tube model for quark mass $m = 3.5$.}
\label{fig:ftqk_en_35}
% FIGURE_FILE qe_vs_beta_35.ax = fig07.ax
}
\figure{
 \caption{Induced charge {\it vs} temperature in the flux tube model.}
\label{fig:ftqtot}
% FIGURE_FILE qtot_vs_T.combo.ax = fig08.ax
}
\figure{
 \caption{Lightest meson-like and baryon-like clusters (two types)
 attached to a fixed quark (burst) in the flux tube model.}
\label{fig:ftlightest}
% FIGURE_FILE lightest.ax = fig09.ax
}
\figure{
 \caption{Typical flux tube lattice at $T \approx 0.78\ T_c$.  The
 origin, indicated by the burst, has been displaced for clarity.}
\label{fig:ftsample_lattice}
% FIGURE_FILE lat24view2.ax = fig10.ax
}
\figure{
 \caption{Flux tube model with $m = 1.0$. Induced quark number density
 {\it vs} distance from the origin. The legend $Q(r\le 3)$ gives the
 integral out to $r = 3$.}
\label{fig:ftdensityvsr}
% FIGURE_FILE combo10.r.ax = fig11.ax
}
\figure{
 \caption{Mean size of cluster connected to the origin {\it vs} inverse
 temperature for the flux tube model.  The errors are smaller than the
 plot symbols.
 }
\label{fig:ftclustersizevsT}
% FIGURE_FILE n0_vs_T.combo.ax = fig12.ax
}
%
%%%%%%%%%%%%%%%%%%%%%%%%%%%%%%%%%%%%%%%%%%%%%%%%%%%%%%%%%%%%%%%%%%%%%%%%%%%%
%   Tables
%%%%%%%%%%%%%%%%%%%%%%%%%%%%%%%%%%%%%%%%%%%%%%%%%%%%%%%%%%%%%%%%%%%%%%%%%%%%
\narrowtext
\begin{table}
\caption{QCD simulation sample size (molecular dynamics time units)
\label{tab:simsize}
}
\begin{tabular}{llr}
$\beta$ & $N_t$ & time  \\
5.15    & 4     & 994.5 \\
5.22    & 4     & 856   \\
5.25    & 4     & 430 \\
5.29    & 4     & 500 \\
5.445   & 4     & 611 \\
5.445   & 6     & 1141 \\
5.445   & 8     & 2665
\end{tabular}
\end{table}
%
%%%%%%%%%%%%%%%%%%%%%%%%%%%%%%%%%%%%%%%%%%%%%%%%%%%%%%%%%%%%%%%%%%%%%%%%%%%%
\widetext
\begin{table}
\caption{\label{tab:scalar5445} QCD simulation results for $6/g^2 = 5.445$ }
\begin{tabular}{lllllllll}
& & & & & \multicolumn{2}{c} {static} & \multicolumn{2}{c} {slice} \\
$N_t$&$T/T_c$&$Q$&${\rm Re} P$& $\left\langle\bar\psi\psi\right\rangle$
                  & $\chi_{\rm s}$&$\chi_{\rm ns}$ & $\chi_{\rm s}$&$\chi_{\rm
ns}$ \\
\hline
4 & 1.5  & -0.380(100) & 0.6480(20) & 0.0768(5) & 0.233(10)& 0.2410(80) &
0.240(20)   & 0.239(13) \\
6 & 1.0  & -0.130(60)  & 0.1070(40) & 0.3150(30)& 0.030(7) & 0.0360(40) &
0.035(16)   & 0.042(13) \\
8 & 0.75 & -0.006(3)   & 0.0065(5)  & 0.1860(5) & 0.009(5) & 0.0019(6)  &
0.010(12)   & 0.008(10) \\
\end{tabular}
\end{table}
%%%%%%%%%%%%%%%%%%%%%%%%%%%%%%%%%%%%%%%%%%%%%%%%%%%%%%%%%%%%%%%%%%%%%%%%%%%%
\begin{table}
\caption{\label{tab:scalarnt4} QCD simulation results for $N_t = 4$}
\begin{tabular}{llllllll}
& & & & \multicolumn{2}{c} {static} & \multicolumn{2}{c} {slice} \\
$6/g^2$&$Q$&${\rm Re} P$& $\left\langle\bar\psi\psi\right\rangle$
                  & $\chi_{\rm s}$&$\chi_{\rm ns}$ & $\chi_{\rm s}$&$\chi_{\rm
ns}$ \\
\hline
5.15  & -0.790(70) & 0.0486(9)  & 0.4945(10) & 0.000(20)& 0.025(6)  &  0.05(3)
& 0.040(13)  \\
5.22  & -0.670(60) & 0.0680(20) & 0.4477(13) & 0.000(30)& 0.000(1)  &  0.01(2)
& 0.023(15)  \\
5.25  & -0.600(60) & 0.0860(20) & 0.4110(20) & 0.070(20)& 0.045(9)  & -0.01(3)
& 0.039(18)  \\
5.29  & -0.046(6)  & 0.4060(20) & 0.1990(30) &   ---    &  ---      &   ---
& 0.171(14) \\
5.445 & -0.006(3)  & 0.6480(20) & 0.0768(5)  & 0.233(10)& 0.241(8)  &  0.24(2)
& 0.239(13)
\end{tabular}
\end{table}
%%%%%%%%%%%%%%%%%%%%%%%%%%%%%%%%%%%%%%%%%%%%%%%%%%%%%%%%%%%%%%%%%%%%%%%%%%%%
\begin{table}
\caption{\label{tab:ftscalars10} Flux tube model results for $m = 1.0$ }
\begin{tabular}{llllllll}
$\beta$&$T/T_c$&$|n|$&$|\ell|$& $n_{\rm vtx}$ & $Q$     &$\left\langle
B^2\right\rangle$ & $N_0$   \\
\hline
1.0&1.88& 0.42220(20)& 0.42037(9) & 157.550(80)&$\ \ 0.00(20)  $&46.50(60)
&894(2)    \\
1.4&1.34& 0.31693(15)& 0.30492(10)& 105.580(70)&$\ \ 0.00(20)  $&35.00(50)
&676(2)    \\
1.6&1.17& 0.25783(15)& 0.23331(11)& 73.680(80) &$-0.01(15)  $&28.40(40) &430(3)
   \\
1.8&1.04& 0.18532(15)& 0.14651(12)& 39.080(60) &$\ \ 0.03(13)  $&19.50(30)
&73.5(1.0) \\
2.0&0.94& 0.10102(14)& 0.06091(10)& 11.980(40) &$-0.44(9)   $&9.53(13)
&7.88(10)  \\
2.2&0.85& 0.04315(10)& 0.01909(5) & 2.490(20)  &$-0.51(5)   $&2.75(4)
&3.25(3)   \\
2.4&0.78& 0.01957(7) & 0.00714(3) & 0.636(8) &$-0.69(3) $&0.834(15)&2.26(2)
\end{tabular}
\end{table}
%%%%%%%%%%%%%%%%%%%%%%%%%%%%%%%%%%%%%%%%%%%%%%%%%%%%%%%%%%%%%%%%%%%%%%%%%%%%
\begin{table}
\caption{\label{tab:ftscalars35} Flux tube model results for $m = 3.5$ }
\begin{tabular}{llllllll}
$\beta$&$T/T_c$&$|n|$&$|\ell|$ &$n_{\rm vtx}$& $Q$      &$\left\langle
B^2\right\rangle$&$N_0$\\
\hline
1.40&1.19 & 0.0530(8)& 0.2687(4) &71.1(3)&$\ \ 0.10(20)  $&1.77(6)& 650(3)\\
1.50&1.11 & 0.0427(7)& 0.2149(5) &51.3(2)&$\ \ 0.00(11)  $&1.35(7)& 535(3)\\
1.60&1.04 & 0.0326(7)& 0.1386(15)&27.9(4)&$\ \ 0.20(10)  $&1.09(7)& 337(4)\\
1.62&1.03 & 0.0303(6)& 0.1202(15)&22.7(4)&$\ \ 0.00(20)  $&0.97(6)& 289(5)\\
1.64&1.01 & 0.0282(7)& 0.0950(20)&16.5(5)&$\ \ 0.20(20)  $&0.94(6)& 206(7)\\
1.66&1.00 & 0.0242(5)& 0.0670(20)&10.1(4)&$-0.30(15)  $&0.71(5)& 117(5)\\
1.68&0.99 & 0.0183(7)& 0.0450(20)&5.8(5) &$-0.17(9)   $&0.54(5)& 59(6) \\
1.70&0.98 & 0.0169(4)& 0.0341(11)&3.8(2) &$-0.26(9)   $&0.42(3)& 35(3) \\
1.80&0.92 & 0.0099(3)& 0.0137(3) &0.89(5)&$-0.75(7)   $&0.12(3)& 8.2(4)
\end{tabular}
\end{table}
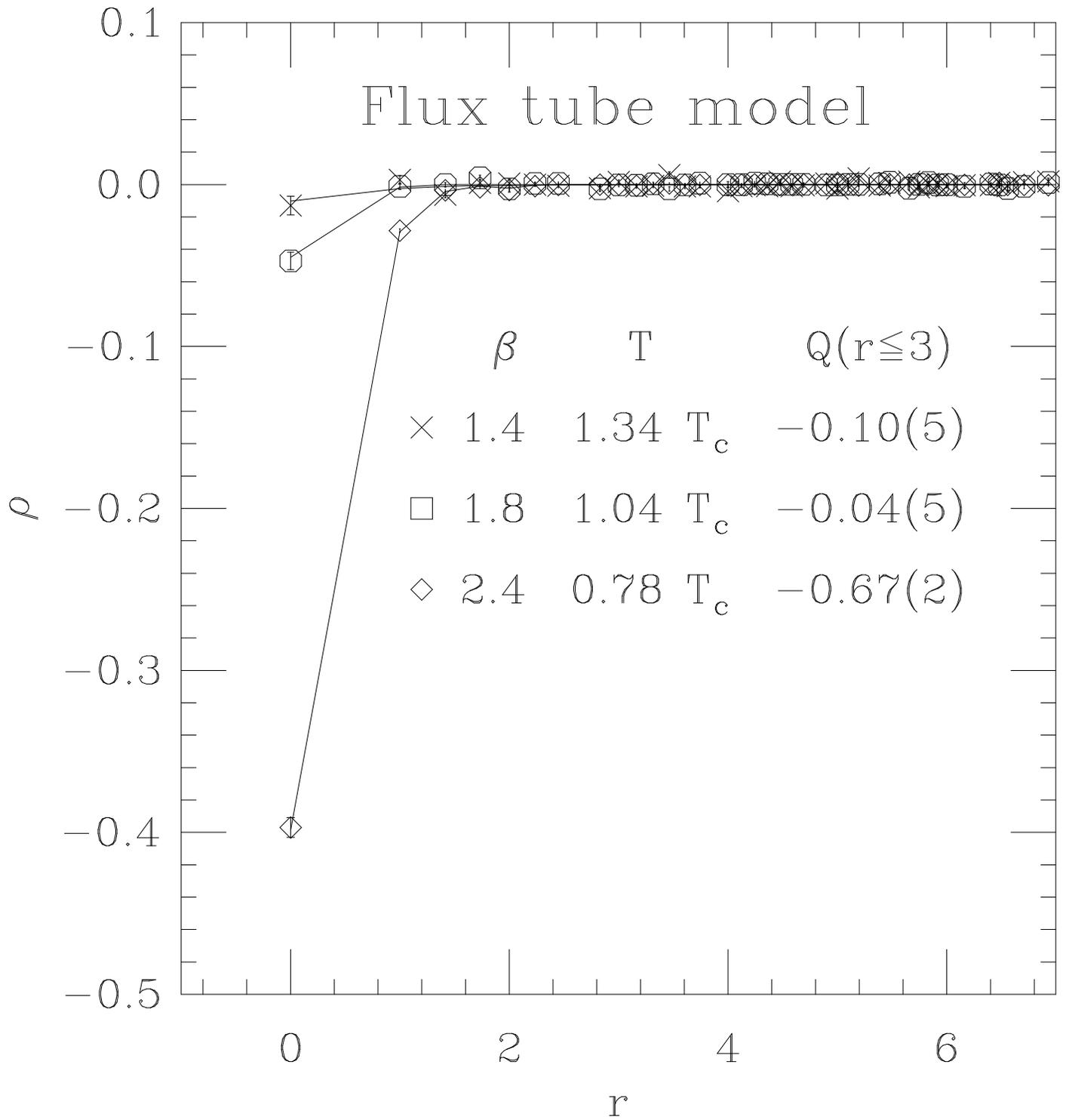
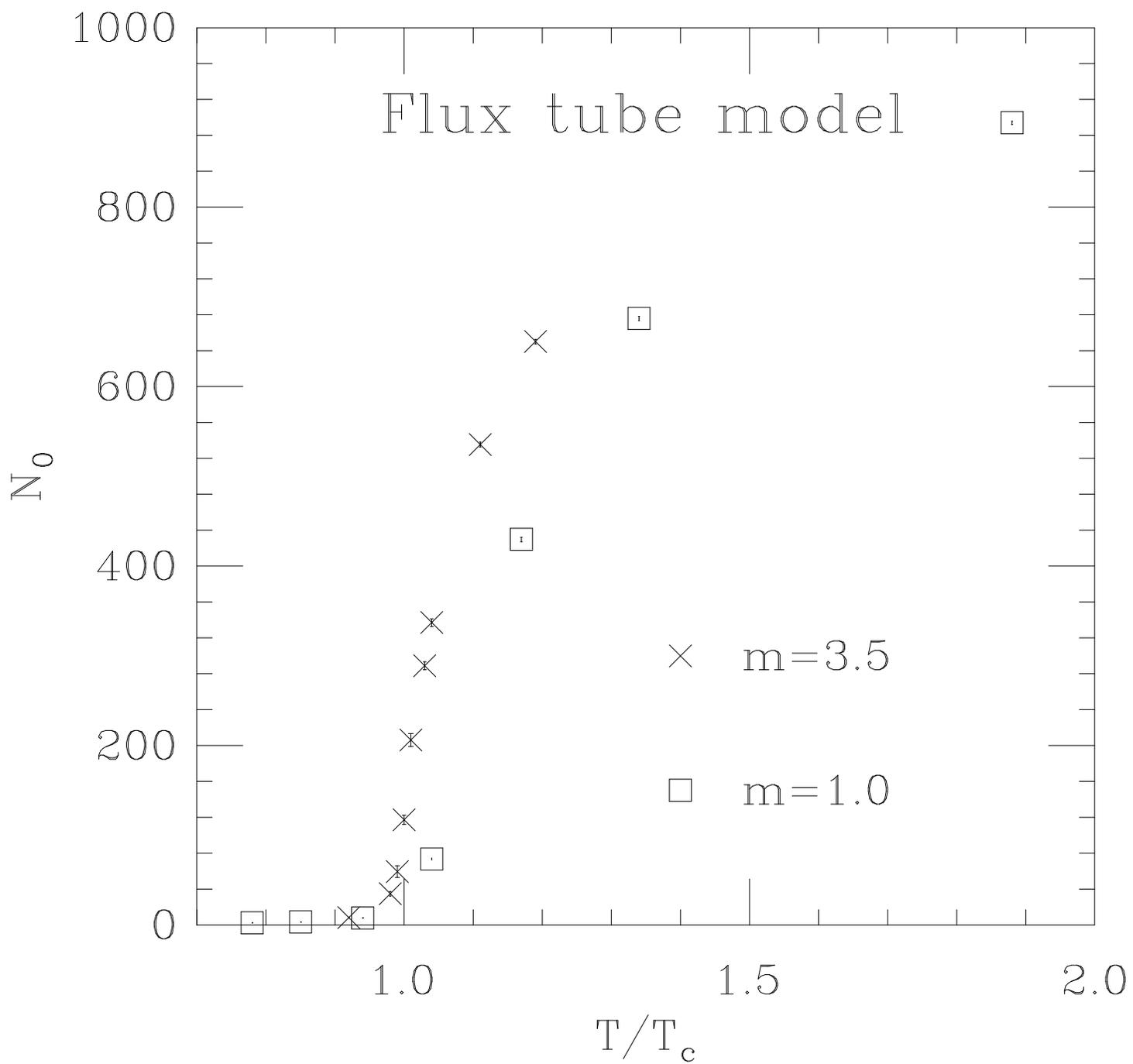
\end{document}